# Causal Wave Mechanics and the Advent of Complexity.
# V. Quantum field mechanics

A.P. Kirilyuk*

Institute of Metal Physics, Kiev, Ukraine 252142

ABSTRACT. The physical consequences of the analysis performed in Parts I-IV are summarised within a tentative scheme of the complete quantum (wave) mechanics called quantum field mechanics and completing the original ideas of Louis de Broglie by the dynamic complexity concept. The total picture includes the formally complete description at the level of the 'average' wave function of Schrödinger type that shows dynamically chaotic behaviour in the form of either quantum chaos (Parts I-III), or quantum measurement (Part IV) with causal indeterminacy and wave reduction. This level is only an approximation, though rather perfect and often sufficient, to a lower (and actually the lowest accessible) level of complexity containing the causally complete version of the unreduced, essentially nonlinear "double solution" proposed by Louis de Broglie. The extended 'double solution with chaos' describes the state of a nonlinear material field and includes the unstable high-intensity "hump" moving chaotically within the embedding smooth wave (quant-ph/9902015,16). The involvement of chaos causally understood within the same concept of dynamic complexity (multivaluedness) provides, at this lower level, de Broglie's "hidden thermodynamics" now, however, without the necessity for any real "hidden thermostat" at a mysterious "subquantum level" of reality. The chaotic reduction of the "piloting" Schrödinger wave, at the higher level of complexity, conforms with the detailed 'wandering' of the virtual soliton. The proposed dynamic multivaluedness (redundance) paradigm serves as the basis for a self-consistent hierarchic picture of the world characterised by a (high) non-zero complexity (and thus irreducible unpredictability), where the complete extension of quantum mechanics is causally interpreted as several lowest levels of complexity.

NOTE ON NUMERATION OF ITEMS. We use the unified system of consecutive numbers for formulas, sections, and figures (but *not* for literature references) throughout the full work, Parts I-V. If a reference to an item is made outside its "home" part of the work, the Roman number of this home part is added to the consecutive number: 'eq. (17)' and 'eq. (17.II)' refer to the same, uniquely defined equation, but in the second case we know in addition that it is introduced in Part II of the work.

---

*Address for correspondence: Post Box 115, Kiev - 30, Ukraine 252030.
 E-mail address: kiril@metfiz.freenet.kiev.ua

# 10. Quantum field mechanics: causal complexity = completeness

As we have already noticed in Part IV, the development of quantum mechanics (involving, in fact, the problems of quantum field theory) has reached today the important turning point. The latter is characterised by the well-pronounced general 'saturation' of the explicative possibilities of various approaches, which even leads to a kind of 'common thinking', despite all the existing distinctions and unresolved problems (see e. g. [1]). It becomes, however, more and more evident that whatever the efforts within this way of thinking, one can hardly hope to arrive at the truly consistent solutions for the fundamental problems ranging from the incompleteness of quantum mechanics and the unified field theory to the introduction of a universal notion of complexity into the fundamental physical theories.

In quantum mechanics (and eventually in any other field) one may discern a more formal approach leading to various, often sophisticated, "interpretations" of the existing concepts, and a more intuitive search for basically simple physical understanding of the objective reality which should inevitably be based on new fundamental concepts. The first direction, stemming from the philosophical position of Niels Bohr, produced a remarkable practical success in explication of the observed phenomena. However, recent years reveal the more and more evident fundamental limits to this kind of development involving the foundations of quantum mechanics, dynamical randomness appearance in quantum world, and further progress in quantum field theory. The second approach, vigorously advocated by Louis de Broglie and seemed to be physically much more justified and qualitatively attractive, has met enormous technical difficulties during the attempts of its practical realisation. Both directions suffered most of all from the well-known "incomprehensible" quantum paradoxes, the **X**-mysteries [2], involving two basic manifestations of the wave-particle dualism, quantum indeterminacy and wave reduction.

In Part IV we have proposed a causal theory of quantum measurement which provides a natural explanation for these two phenomena based on the postulate of the fundamental dynamic uncertainty, the same one that appears in the self-consistent description of quantum chaos in Hamiltonian systems, Parts I-III. This causal interpretation of the **X**-mysteries, besides the proposed solution itself, permits one to effectively separate them from other problems of quantum theory, the **Z**-mysteries [2], which are much less puzzling, related rather to technical difficulties of formal description, but are, in principle, comprehensible.

In terms of de Broglie picture it means that there still remains to find a particular formulation of the (material) nonlinear wave dynamics giving soliton-like solutions and compatible, at the same time, with the quasi-linear Schrödinger formalism. But there is no serious doubts that it is possible in principle and that many nonlinear equations can provide chaotic soliton-like structures approaching the desired properties. In addition to the pioneering works of de Broglie (see [3-6] and the references therein), we can cite refs. [7,8]

as providing recent examples of particular nonlinear particle-like solutions of this type (see also the footnote below). What is important is that now, provided with the fundamental causal origins of uncertainty and reduction that complete essentially the causal interpretation in general, we can 'calmly' make our choice among those nonlinear equations avoiding the confusing confrontation with such problems as the mysterious "hidden thermostat" or parasitic "empty wave". It is easy to see that independent of their particular form, similar problems will always be present in any version of quantum mechanics, unless it proposes some really fundamental and universal possibility of combining complexity with the starting quasi-linear quantum formalism (in our approach it is reduced to the postulate of the fundamental dynamic uncertainty). We are going to show now that this decisive advantage of restored causality of the wave-particle dualism can be used to construct a tentative general scheme of the eventual complete quantum mechanics/quantum field theory which we call *quantum field mechanics*. We shall see that this theory is nothing but a natural continuation of de Broglie's ideas complemented with the concept, and the formalism, of the fundamental dynamic uncertainty.

We start the description of quantum field mechanics with its axiomatic structure presented by two basic concepts. The first one states that the physical entity forming the basis of the World is an effectively nonlinear material field (wave) producing unstable, but always present, soliton-like localised structures, or 'particles'. This field and the particles it produces obey (effectively) nonlinear equations to be found, but at the same time their behaviour is compatible with the Schrödinger equation (in general, it is the modified Schrödinger equation, see section 5.III).[*] This nonlinear quasi-particle may be embedded in the much more extended accompanying field, even though this should be finally confirmed within the detailed theory. There should eventually exist several different types of such nonlinear field corresponding to the elementary particles, but they all stem from the same origin, the nonlinear 'proto-field', splitted into a number of components presumably by the same fundamental mechanism of FMDF (fundamental multivaluedness of dynamical functions). We call the above group of statements the *wave postulate*. In fact, it simply provides an answer to the question 'what?' (i. e. what is it that exists as the irreducible physical basis of the World?) and should certainly be completed by the precise nonlinear formalism compatible with the Schrödinger equation and the corresponding basic experimental facts.

The second postulate states that the behaviour of these nonlinear fields is compatible with the concept of the fundamental dynamic uncertainty, i. e. it is generally chaotic, in certain well-defined sense. The necessity of this second axiom follows from the description of complex quantum dynamics presented in Parts I-III and re-established in Part IV for the measurement process. Indeed, we have seen that even the Schrödinger formalism of quantum mechanics can provide, in a self-consistent manner, the true dynamical chaos, complexity, and causal quantum indeterminacy and reduction, on condition that we accept certain modified, or 'effective', form of dynamic equations which is deduced from the ordinary form and comprises all its solutions, but contains also many additional solutions. The nonlinear quasi-particle structures from the first postulate should certainly demonstrate complex behaviour of the same fundamental origin. This is necessary already in order to satisfy the demand of consistency with the modified Schrödinger equation, but also to properly explain the particle behaviour itself. In fact, as we have seen from the preceding analysis, complexity appears inevitably for practically any system with nontrivial structure/interactions, which is certainly the case for the introduced nonlinear field. In summary, this second basic statement simply fixes the choice for the modified formalism with its dynamic multivaluedness, as opposed to the ordinary 'single-valued' formalism, and will be called the *dynamic complexity postulate*. It answers the second primary question, 'how?' (i. e. how does it behave, the nonlinear field?). Note that it is the particular answer proposed that opens the way for the self-consistent introduction, within the first postulate, of a *material* wave, rather than a "wave of probability density" (because we have *obtained* the probability in a causal way).

---

[*] The simplest, apparently noncontroversial and universal enough possibility is that it is our modified Schrödinger formalism itself that directly determines the dynamics of the chaotic soliton within the effectively nonlinear Schrödinger wave field. The detailed development and eventual confirmation of this particular solution deserves, however, separate publication(s) (see e-prints quant-ph/9902015,16 in Los-Alamos archives).

It is the complexity postulate which substantially amplifies the basic propositions of de Broglie approach (the latter entering, in fact, the wave postulate above) and considerably facilitates its best formulation and realisation. Indeed, if we apply the universal formalism of FMDF to the anticipated soliton-like structures we shall reveal their chaotic behaviour reduced to their permanent quasi-random motions within the embedding "pilot-wave". By the way, it corresponds well to the known instability of the most of soliton-like solutions of nonlinear equations; we shall obtain a kind of 'virtual soliton' which moves by constantly disappearing and reappearing at different positions in a random fashion (cf. multiple realisations of a system, in our approach). This does not preclude the existence of certain quasi-regular part of its behaviour governed, in particular, by the interaction with external objects, as it is predicted by the FMDF concept. This type of complex behaviour of the soliton-like core of elementary quantum object is quite useful, and even necessary, for the causal explanation of the uncertainty and reduction during quantum measurement. That could explain why the idea about such random motion was introduced by de Broglie into his concept, in the later period, under the name of "hidden thermodynamics" of a particle ("thermodynamique cachée") [9] (see also [10]). One cannot escape the surprise of the precise physical similarity between this prophetic idea and what has started to emerge much later under the name of dynamical chaos. However, at the time of this assumption de Broglie was forced to postulate the physical existence of the corresponding "hidden thermostat" serving as a source of random driving force for the hidden thermodynamics. As we have seen above, without the FMDF concept one could not avoid this assumption even today because the conventional quantum formalism cannot provide any intrinsic dynamical randomness. One may say thus that the role of the dynamic complexity postulate is to provide the "hidden thermodynamics" without any real "hidden thermostat", the latter complicating unnecessarily the construction of the complete quantum theory.[*)] These implications of the second postulate of quantum field mechanics form the *third level* of chaos involvement with the foundations of quantum mechanics (the first two levels are specified in section 8.IV).

The role of complexity in quantum field mechanics is, however, yet more involved. Indeed, there is the other part of the double solution, the Schrödinger wave function. This irreducible dual partner of the soliton-like 'particle' is in fact as real as the latter because it gives easily observable effects like diffraction. Of course, physically its existence may seem to be less transparent than that of the isolated localised 'particle', but from the other hand the precise mathematical formalism, providing experimentally confirmed results, exists at present just for this nonlocal quasi-linear part of solution. It is because of this latter circumstance that we were able to obtain directly the quantum-mechanical indeterminacy and wave reduction for this part of the double solution, even though this has demanded a non-trivial involvement of the postulate, and the formalism, of the fundamental dynamic uncertainty. We argue, however, that the two versions of the chaotic behaviour of the double solution, corresponding to its two components, represent the same unity as those two components themselves. One deals here with two dualistic descriptions of the *same* indivisible object and its complex behaviour. In particular, the discovered causal randomness in the (modified) Schrödinger wave behaviour accompanied by its reduction (localisation) physically correspond precisely to chaotic wandering and localisation of the virtual soliton in the process of its interaction with the quasi-solitons of the instrument. This physical correspondence is ensured and expressed by the universal nature of our basic concept, and the formalism, of the fundamental multivaluedness which can be applied at each level of the description. We have been able to partly confirm this profound agreement above by showing that the localised singularity can be naturally incorporated in our results on incoherent reduction. The fundamental involvement of local excitations in the measurement process (section 9.1.IV) also seems to be qualitatively reducible to virtual soliton interactions. It is clear that a much more reliable support for this correspondence between the two components of the dualistic description can be obtained only within the detailed nonlinear

---

[*)] One can even trace more detailed relations with the results of de Broglie's theory. In particular, it seems to be rather evident that the entropy of the isolated particle, introduced by de Broglie to account for the internal Brownian-like motion of the 'hump' [9,10], is nothing but the entropy of the chaotic dynamics or, in terms of our general analysis, its complexity determined by the number of realisations for a system (see section 6.III). Similarly, de Broglie's internal temperature of the isolated particle can be associated with the rate of realisation change depending, in particular, on the nonlinearity of a system (section 3.II). Once the nonlinear double solution equations are found, our basic FMDF method will provide explicit expressions of those parameters of the "hidden thermodynamics" in terms of the most fundamental characteristics of the elementary field (see the book presented by e-print physics/9806002). Here we can add only that this interpretation is quite consistent with the efforts of de Broglie to relate the internal thermodynamics of a particle to its 'external', global dynamics (see [10]).

formalism providing both parts of the double solution as well as the averaged Schrödinger description.

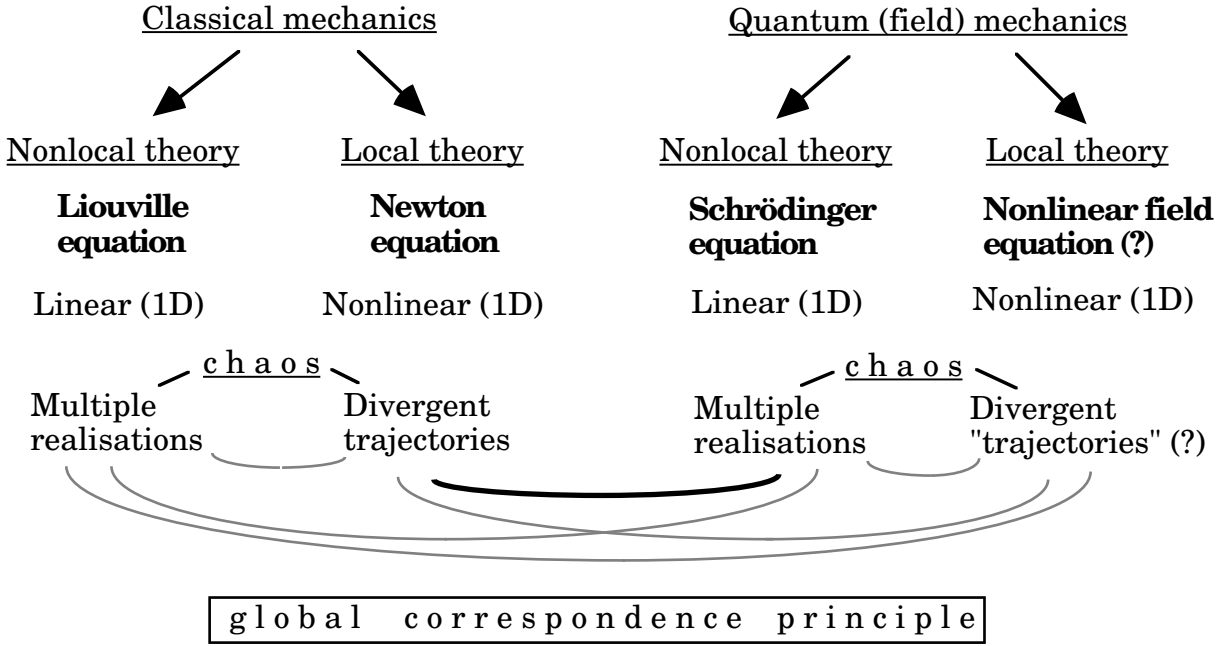

**Fig. 5.** Universal dualistic structure of physical theories exemplified by classical mechanics and quantum field mechanics: local vs. nonlocal description including complex (chaotic) behaviour.

One can better see the general logic of this interplay between the components of the double solution and the fundamental dynamic uncertainty, within the quantum field mechanics, with the help of a scheme representing the dualistic structure of physical theories, Fig. 5. We suppose that the introduced subdivision of each field into two dualistic versions, based respectively on the local and the nonlocal approaches, can eventually be specified for every field of physics large enough, but here we restrict ourselves only to the pertinent structure of classical mechanics and quantum field mechanics. We see that in each case the local and the nonlocal approaches are characterised by the same repeated features: nonlocal theory is expressed by the initially linear (e. g. for one degree of freedom) formalism providing chaos (and the effective nonlinearity) by the mechanism of FMDF, whereas the local approach is presented by a nonlinear starting formalism and gives chaos typically in the form of the divergent trajectories, or their analogues.[*)] We emphasize that theories of both types are *formally* complete (this completeness can be realised, in particular, due to the introduced universal dynamic uncertainty) and thus generally equivalent among them, but at the same time they are *physically* complementary

---

[*)] The manifestation of chaos within the local theory can also be described as the effective nonlinearity using the method of FMDF, due to universality of the latter (see section 6.III), but we shall not specify the details here.

to each other, and thus each of them is physically incomplete. Usually the local approach seems to be intuitively more comprehensible (and thus attractive), but it does not provide the global dynamical picture, whereas the nonlocal one hides many particular details of an extended system, but permits one to understand its dynamics as a whole. In accord with this structure of physical theories, multiple correspondence connections can be established between equivalent approaches from different fields (they are shown in the bottom part of Fig. 5 for the considered case of classical mechanics and quantum field mechanics). In particular, in part II we have established the quantum-classical correspondence between the nonlocal quantum and the local classical descriptions (solid line in Fig. 5). This does not diminish the interest to specify other possible correspondence connections between different fields (dotted lines), even though some of them will first demand the completion of the respective approaches themselves. We call all the self-consistent system of these connections between the complete classical and quantum theories the *global correspondence principle*. Its full realisation will demand considerable additional developments both in quantum and in classical description of complex dynamical behaviour. Nevertheless, already at the present stage the whole picture seems to be rather consistent and transparent, at least qualitatively. In particular, we have completed, in principle, the nonlocal part of quantum field mechanics, which can help, as we have explained above, to advance with more confidence while constructing the local part. We see once more that quantum field mechanics can be considered as the direct extension of de Broglie ideas (the wave postulate) complemented with the postulate of dynamic complexity.

It can be interesting also to consider the relation between the quantum field mechanics and the Einstein approach to quantum mechanics and 'unified' quantum field theory. We have seen that in our general scheme the nonlocal part, corresponding to the quasi-linear quantum mechanics of Schrödinger, should be deducible from the full nonlinear description representing the nonlinear quantum field theory. The contemporary quantum mechanics is thus rather a derivative theory starting from the 'really fundamental' nonlinear field mechanics, the relation that was generally anticipated by Einstein. Note, however, that according to the scheme of Fig. 5, the nonlocal part is equivalent to the local one, so that finally one obtains the most complete theory, quantum field mechanics, rather as the involved combination of "quantum mechanics" and "quantum field theory". This is related to a fundamental distinction from the Einstein approach: whereas he tried to advance by deduction starting from the most general 'mathematical', purely mechanistic formulation of the 'unified theory' leaving no place to chance ("God does not play dice"), we inherit from de Broglie the analysis by induction starting from the lowest, 'physically' interpreted levels, where the causally emerging 'chance' plays the crucial role. Of course, both the local and nonlocal approaches should give eventually the same causal description, 'meeting' somewhere between the general field equations (properly interpreted within the intrinsically multivalued, nonunitary description) and the nonlinear equations for soliton-like structures compatible with the (modified) Schrödinger equation. Within de Broglie's strategy accepted in quantum field mechanics, one may expect to obtain first a self-consistent dualistic description for one kind of elementary particle (nonlinear field) and then to generalise the results to other field-particles, whereas within the Einstein program one tries to obtain the coupled equations for a group of fields (or even all of them) and then 'descend' down to splitting into elementary fields and their 'concentration' into particles, but the origin of the wave-particle duality and related indeterminacy remain basically unclear because of the dominating irreducible unitarity. The universal character of FMDF mechanism supposes that in reality any 'splitting' into new entities should occur as a result of the fundamental multivaluedness (redundance) of solutions of a hypothetical unified field equation and therefore cannot be separated from the intrinsic causal randomness.

In this connection, we can recall also the well-known general conflict between Einstein's vision of totally regular reality and the probabilistic elements of quantum mechanics (in fact, he was opposed to *any* idea about the *basic, irreducible* randomness of the World). The incompleteness of the standard quantum mechanics has been acknowledged by everybody, but Einsteinian kind of 'understanding', closely related to the imposed absolute power of mathematical 'symbolism' in the canonical science, naturally sees any *unpredictable*, probabilistic randomness as a violation of causality *in itself*, without even asking for any its causal origin (this explains also why Einsteinian vigorous objections against quantum mechanics could always 'peacefully' coexist with the absence of any truly complete, physically sound basis behind the purely abstract substantiation of his relativity; see e-print quant-ph/9902016 for more detail). Our results open the unique way for resolution of this basic conflict of the unitary science: we propose an *intrinsically complete (consistent), universal* source of randomness, the fundamental dynamic uncertainty, which, however, *reduces* it to a *dynamical* effect, the dynamic multivaluedness of realisations, that is *causally deduced* from the *deterministic* equations. This can be interpreted as a new *definition* of the true randomness realising the unique agreement with both its experimental manifestations and philosophical consistency of the total world's picture (cf. section 6.III).

It is not out of place to mention that our results, and especially those directly reflecting the fundamental dynamic uncertainty, can be seen as a generalisation, or a more consistent version, of a number of known *interpretations* of quantum mechanics (see e. g. [2,11-13]). We shall not repeat the discussion of the evident connections with the Copenhagen and pilot-wave interpretations. The most interesting is the relation to the many-worlds interpretation [14] stating that the World as a whole is splitted into multiple branches during each quantum measurement under each object, these branches corresponding to the plurality of possible issues of measurement forming quantum indeterminacy. In our results the latter is indeed involved with splitting, but this relation is rather deduced than introduced artificially, and what is especially important, it concerns only the particular system of object and instrument for a given measurement process, and not the whole World. Moreover, our splitting into many realisations does not mean the real 'parallel' coexistence of all those realisations. What really exists is one of the realisations per each run of the measurement process, this realisation being chosen at random, but with the known probability, from their ensemble which can never be represented explicitly by more than one its member. In return, one can calculate and know this ensemble, and the corresponding probabilities, independent of a real, experimentally observed, process of measurement. We arrive thus at a similarity with ensemble, or statistical, group of interpretations of quantum mechanics [15]. The relation consists in the fact that our fundamental multivaluedness provides, in fact, the causal physical source of the necessary plurality of ensemble members which otherwise should be directly postulated. This reflects the objective and inevitable appearance of stochasticity in quantum mechanics, but only the discovered true deterministic randomness of quantum systems can ensure its universal fundamental origin. And finally, the more recent "quantum-trajectories" interpretation [16], related to Feynman path integral, can be traced in the scheme of quantum field mechanics in the causally extended form of chaotic motion of the *real* virtual soliton-particle within the embedding field. In section 9.1.IV we have seen also that even the exotic ideas about the irreducible subjective influence of a conscious observer on the measurement process [17] can find their quite objective causal counterpart in our description.[*]

We conclude this section with the emphasis on the role of complexity, the latter being always understood in the same well-defined sense (see eqs. (34), section 6.III), in the proposed causal explanation for quantum indeterminacy and wave reduction. We argue that this implication of deterministic randomness in the resolution of the most puzzling quantum **X**-mysteries is inevitable for at least two complementary reasons. First, the resolution of such basic problems of the wave behaviour could only be possible at the expense of a new concept, not less fundamental than the wave postulate itself (see also [2]), and it is not easy to imagine another candidate for this role, apart from a universal postulate introducing dynamic complexity. Second, the dynamic chaos itself should certainly find its place in quantum mechanics describing the complex world, and this place can only be one of the basic ones. The fact that up to now the true chaos seemed to be incompatible with quantum mechanics just shows, as we have seen, that the standard quantum formalism is not adapted to interpretation of complexity. It is, by the way, the case of any nonlocal approach including that of Liouville equation in classical mechanics (see Fig. 5). We have demonstrated, however, that already simple algebraic transformations of the ordinary formalism lead to the appearance of the fundamental dynamic uncertainty that can manifest itself as the 'ordinary' quantum chaos ('fictitious measurement', uncertainty of the object) or as the fundamental quantum indeterminacy (real measurement, uncertainty of the instrument). From the other hand, the same fundamental quantum indeterminacy should become evident, and intuitively more transparent, in the anticipated local formulation of quantum field mechanics, even though this demands construction of a local nonlinear formalism, in continuation of de Broglie ideas. Now, however, the latter task seems to be much more feasible: in parts I-V we have revealed different versions of a universal mechanism showing how the *effective nonlinearity*, being a synonym of complexity (see especially sections 6.III and 9.2.IV) and really existing in the physical world, can be put into a natural explicit form just by properly presented formalism, without any artificial additions. Once appeared after the long and vain search for it, this true and realistic nonlinearity of wave mechanics will certainly give us a variety of the known, anticipated, and now inconceivable possibilities.

---

[*] One particular addition to this brief excursus into the 'science of interpretations' concerns the canonical objection against the 'hidden-parameter' type of approach in quantum mechanics stating that it should inevitably imply the existence of infinitely rapid motions within the wave field. Our causal reduction is not subjected to these difficulties: the nonlinear elementary field represents a whole indivisible object at any stage of its evolution, and no experiment can 'trace' the individual motions of the virtual 'hump' or other parts of the wave; the field can 'freely choose' the centre of its shrinking among many possible ones, but it can never be reduced to several different centres within the same action of reduction. Any 'infinitely rapid motions' within the measured wave cannot be associated thus with propagation of a signal measurable in at least two different points, and the relativistic limitations do not apply.

It would not be out of place to recall that the founders of the Copenhagen interpretation, led by Niels Bohr, had seen the final victory of their approach in the definite exclusion from quantum physics of the ordinary 'macroscopic' intuition, based on everyday experience and especially on its 'ordinary' human logic. Now, seventy years after they have won, it is precisely this type of logic that reappears as the non-contradictory causal scheme of wave mechanics within the described synthesis between the "defeated" intuitive approach of de Broglie and the universal concept of dynamical complexity. We have seen that almost humanly intricate, unpredictable and multiform, behaviour of the effectively nonlinear material wave becomes quite natural, and even inevitable, provided a 'gentle', logically correct modification of the basic formalism is accepted in exchange for the irreducible complexity of the world. This 'humanization' of quantum mechanics has been expected as one of the necessary constituents of the beginning, and unavoidable, fundamental return of wholeness into the entire system of knowledge.

The involvement of complexity at the very basis of quantum mechanics is profoundly related also to the universal hierarchical structure of the World. Indeed, the most fundamental level of description of the complex world should certainly contain dynamic complexity in explicit form. This demand is now satisfied for quantum mechanics within the FMDF formalism. From the other hand, one should be able to obtain complexity at any higher level of description, e. g. in classical mechanics, in distributed system behaviour, etc., without leaving that level. It is extremely important, that it is the *same* mechanism of the fundamental dynamic uncertainty that provides complexity (chaos) at each level (see section 6.III), though with some specific details characteristic of that level and concerning rather the form of the results.[*)] It means that we have the *double correspondence* between different levels of complexity: the *direct* one, where the results at a higher level can be deduced from the more fundamental description (e. g. the results for classical chaotic systems can be obtained within the purely quantum-mechanical consideration, see sections 2.3.II, 3.II); and the *conceptual* correspondence, where the complexity at a higher level can be obtained without any reference to the underlying more fundamental levels, but within the same concept and method as the ones that reveal complexity at lower levels. This 'vertical' double correspondence, accompanied with the 'horizontal' global correspondence principle in the sense of Fig. 5, provides another evidence in favour of a self-consistent holistic picture of the Complex World outlined throughout the present work.

We can give finally a well-substantiated positive answer to the basic questions (35) considerably extending our preliminary answers (36), (89) and outlining a feasible issue towards the physically and formally complete quantum field mechanics:

$$\begin{array}{r} \text{Quantum mechanics (in the modified form) obeys} \\ \text{the (global) correspondence principle.} \\ \text{It is formally complete, but physically incomplete.} \\ \text{It can be extended by addition} \\ \text{of the local effectively nonlinear theory.} \end{array} \quad (90)$$

---

[*)] In particular, and this is symptomatic, the formalism applied above to reveal the manifestation of dynamic uncertainty in the measurement process (section 9.1.IV) can be used with only minor changes for the description of complex behaviour in dynamic systems from a very large class. To obtain such general description it is sufficient, in fact, to consider the measured object and the instrument as abstract interacting physical systems characterised by their states and the respective operators that can eventually be specified for each particular problem.

# ACKNOWLEDGEMENTS

The author gratefully acknowledges the support of the Société de Secours des Amis des Sciences (Paris) during a large part of this work.

He is indebted to Dr. Georges Lochak, the Director of the Fondation Louis de Broglie, for the valuable scientific discussions, indispensable all-round help and attention. In particular, the ideas of parts IV,V have been largely inspired by the activity of the Foundation including both the particular results and the general scientific approach continuing those of de Broglie.

The discussion with S. Weigert from Universität Basel was useful for clarification of general issues concerning quantum chaos.

The completion of this work has only been possible due to financial and moral support from my family.